Increase in $T_c$ and change of crystal structure by high-pressure annealing in $BiS_2$-based superconductor $CeO_{0.3}F_{0.7}BiS_2$

Joe Kajitani[1], Takafumi Hiroi[1], Atsushi Omachi[1], Osuke Miura[1], and Yoshikazu Mizuguchi[1]*

1. Department of Electrical and Electronic Engineering, Tokyo Metropolitan University, 1-1, Minami-osawa, Hachioji, 192-0397, Japan

**Abstract**

Recently, several types of $BiS_2$-based superconductor such as $Bi_4O_4S_3$, $REO_{1-x}F_xBiS_2$ (RE: rare earth) and $Sr_{1-x}La_xFBiS_2$ have been discovered. In this study, we have investigated the crystal structure and the superconducting properties for two kinds of polycrystalline samples (As-grown and high-pressure-annealed samples) of the $BiS_2$-based superconductor $CeO_{0.3}F_{0.7}BiS_2$. We found that both the As-grown and the high-pressure-annealed $CeO_{0.3}F_{0.7}BiS_2$ samples show bulk superconductivity. A higher $T_c$ was observed in the high-pressure-annealed sample. The $T_c$ of $CeO_{0.3}F_{0.7}BiS_2$ increased from $T_c^{zero}$ = 2.7 K to $T_c^{zero}$ = 3.7 K by high-pressure annealing. The lattice constant of *a* and *c* axis did not show remarkable differences between the As-grown and the high-pressure-annealed samples. Nevertheless, the peak symmetry of the (200) peak seemed to become more symmetric while the (004) peak did not show such a difference, indicating that the crystal structure within the *ab* plane changed to a higher-symmetric phase, a perfect tetragonal, by high-pressure annealing in $CeO_{0.3}F_{0.7}BiS_2$.

## 1. Introduction

Recently, several types of $BiS_2$-based superconductor such as $Bi_4O_4S_3$,[1] $REO_{1-x}F_xBiS_2$ (RE: rare earth), [2-6] and $Sr_{1-x}La_xFBiS_2$ [7,8] have been discovered. The typical parent material $LaOBiS_2$ has a crystal structure composed of an alternate stacking of $BiS_2$ conduction layers and LaO blocking layers. Electron carriers, which are essential for the appearance of superconductivity in the $BiS_2$-based family, can be controlled by manipulating the structure and composition at the blocking layers. In $LaOBiS_2$, electron carriers can be generated by a partial substitution of $O^{2-}$ by $F^-$.[2,9] Electrical resistivity and magnetic susceptibility measurements under high pressure revealed that the transition temperature ($T_c$) of the $BiS_2$-based family was sensitive to application of external pressure and can be significantly increased[11-13] as observed in the Fe-based superconductor family.[14-17]

In addition, superconducting properties of the $BiS_2$-based superconductors strongly depend on the sample preparation method. Solid-state-reacted (As-grown) $LaO_{1-x}F_xBiS_2$ shows filamentary superconductivity: in other words, the superconducting volume fraction is obviously low, whereas superconducting states are evidently generated. To induce bulk superconductivity, high pressure annealing is effective.[2,10] Using the high-pressure technique, the onset of $T_c$ of $LaO_{0.5}F_{0.5}BiS_2$ reaches a value of 11 K. It was suggested that the crystal structure obviously changed after high-pressure annealing. Recently, we revealed that uniaxial lattice contraction generated along the *c* axis was positively linked to the enhancement of $T_c$ in high-pressure-annealed $LaO_{0.5}F_{0.5}BiS_2$ and $PrO_{0.5}F_{0.5}BiS_2$.[18,19]

In this study, we focus on $CeO_{0.3}F_{0.7}BiS_2$. In the previous studies, it was reported that As-grown $CeO_{0.3}F_{0.7}BiS_2$ does not show bulk superconductivity.[3,4] In addition, the magnetic ordering, which was regarded as a ferromagnetic ordering, was observed. Bulk superconductivity is induced by application of high pressure[11] or annealing the As-grown samples under high pressure (high-pressure annealing).[4] Interestingly, it was found that the magnetic ordering at the blocking layer (Ce moment) and bulk superconductivity at the $BiS_2$ layer could coexist. So far, the correlation between the superconducting properties and the crystal structure under high pressure in $CeO_{0.3}F_{0.7}BiS_2$ remains to be clarified. In this article, we have investigated the correlation between the crystal structure and the superconducting properties of the As-grown and the high-pressure-annealed $CeO_{0.3}F_{0.7}BiS_2$ samples.

## 2. Experimental detail

Polycrystalline samples of $CeO_{0.3}F_{0.7}BiS_2$ were prepared by a solid-state reaction using powders of $Bi_2O_3$ (99.9 %), $BiF_3$ (99.9 %), $Bi_2S_3$, $Ce_2S_3$ (99.9 %), and grains of Bi (99.99 %). The $Bi_2S_3$ powder was prepared by reacting Bi (99.99 %) and S (99.99 %) grains in an evacuated quartz tube. Other chemicals used in this study were purchased from Kojundo-Kagaku Laboratory. The starting materials with a nominal composition of $CeO_{0.3}F_{0.7}BiS_2$ were mixed-well, pressed in pellets, sealed in an evacuated quartz tube and heated at 700 ºC for 10 h. The obtained products were ground, sealed into an evacuated quartz tube and heated again under the same heating condition to obtain a homogenized sample; in this article, we call this sample *As-grown* sample. The obtained sample was annealed at about 600 ºC under a high pressure of about 3 GPa for 1 h using a cubic-anvil high-pressure synthesis instrument; in this article, we call this material *HP* sample. All the obtained samples were characterized by X-ray diffraction using the $\theta$–$2\theta$ method with a CuK$\alpha$ radiation. The temperature dependence of electrical resistivity was measured using the four-terminal method. The temperature dependence of magnetization was measured using a superconducting quantum interface device (SQUID) magnetometer with an applied field of 5 Oe after both zero-field cooling (ZFC) and field cooling (FC).

## 3. Results and discussion

Figure 1(a) shows the X-ray diffraction patterns of the As-grown and the HP samples of $CeO_{0.3}F_{0.7}BiS_2$. Almost all of the obtained X-ray peaks were explained using the tetragonal *P*4/*nmm* space group. The numbers displayed in the profile indicate Miller indices. The obtained X-ray profiles are quite similar, but the peaks of the HP sample are slightly broadened. To discuss the change of the lattice constants in detail, the enlarged X-ray profiles around the (004) peaks and the (200) peaks are shown in Figs.

1(b) and 1(c), respectively. To clarify the changes in X-ray peaks, the peak intensities of (102) in Fig. 1(b) and (200) in Fig. 1(c) were normalized to 1, respectively. The (004) and (200) peak position does not show a remarkable differences between the As-grown and the HP samples. The lattice constants estimated from the peak positions are $a$ = 4.0477 Å and $c$ = 13.429 Å for the As-grown sample, and $a$ = 4.0477 Å and $c$ = 13.429 Å for the HP sample. The calculated lattice constants for the As-grown and the HP samples are the same within the resolution of our Laboratory level X-ray diffraction instrument.

Nevertheless, it can be noted that the shape of the (200) peak of the As-grown sample is relatively asymmetric. The (200) peak has a small hump around $2\theta = 45^{\circ}$, which indicates that the *ab*-plane could be strained, and the crystal structure symmetry would be not a perfect tetragonal for the As-grown sample. In contrast, the hump structure disappears in the (200) peak of the HP sample. These facts indicate that the crystal structure changes to a higher-symmetric phase (more perfect tetragonal) by high-pressure annealing in $CeO_{0.3}F_{0.7}BiS_2$. In addition, the (004) peaks for both samples are almost symmetric, which indicates that there is no change in the structure symmetry and/or lattice strain along the $c$ axis.

Figure 2 shows the temperature dependences of electrical resistivity for the As-grown and the HP samples of $CeO_{0.3}F_{0.7}BiS_2$ below 15 K. The As-grown sample shows zero-resistivity states below $T_c^{zero}$ = 2.7 K. The HP sample shows zero-resistivity states below $T_c^{zero}$ = 3.7 K. It is hard to define the onset of $T_c$ in the resistivity measurements because the resistivity for both samples gradually begins to decrease below ~ 8 K. The broad transition is caused by the sensitivity of the superconducting properties to the change in the local crystal structures or strain.

Figure 3 shows the temperature dependence of magnetization for (a) the As-grown and (b) the HP samples of $CeO_{0.3}F_{0.7}BiS_2$. For both samples, magnetic ordering was observed below a magnetic transition temperature of 7.5 K as reported in the previous studies.[3,4] The magnetic transition temperature does not change by high-pressure annealing. The As-grown sample shows the superconducting $T_c^{mag}$ of 2.7 K, while the HP sample shows the superconducting $T_c^{mag}$ of 3.5 K. These results almost correspond to the $T_c^{zero}$ observed in the resistivity measurements for the As-grown and the HP samples. A large shielding volume fraction is observed in the ZFC data at 2 K for both samples, indicating both samples show bulk superconductivity.

The fact that the As-grown sample shows bulk superconductivity seems to be inconsistent with the previous works.[3,4] We consider the difference in the observed properties of $CeO_{0.3}F_{0.7}BiS_2$ between this study and the previous studies was due to the difference of the sample synthesis conditions. In fact, the annealing temperature is different. As shown in the X-ray diffraction part, the changes in the crystal structure after high-pressure annealing are not obvious as compared to the case of $LaO_{0.5}F_{0.5}BiS_2$, in which a large uniaxial contraction along the *c* axis is observed for the HP sample.[18] Therefore, a slight change in the crystal structure could induce bulk superconductivity in $CeO_{0.3}F_{0.7}BiS_2$. Then, bulk superconducting states are realized in the As-grown sample by optimizing the synthesis temperature.

To understand the increase in $T_c$ by high-pressure annealing in $CeO_{0.3}F_{0.7}BiS_2$, we suggest two possibilities. The first scenario is the slight change in the lattice symmetry within the *ab* plane. As shown in Fig. 1(c), the shape of the (200) peak of the As-grown sample is relatively asymmetric, and the (200) peak becomes symmetric by high-pressure annealing. A higher symmetry within the *ab* plane may be important for raising $T_c$ in $CeO_{0.3}F_{0.7}BiS_2$. The other possibility is that the atoms locally moved without any changes in lattice constants by high-pressure annealing. For example, the *z* coordinate of the interlayer S site could easily move or fluctuate because the $S^{2-}$ is combined with not only Bi in the superconducting layer but also Ce in the blocking layer. If this scenario is real, the electronic states which enhance superconducting properties could be modified without any change in lattice constants.

## 4. Conclusion

We investigated the crystal structure and superconducting properties of the As-grown and HP samples of $CeO_{0.3}F_{0.7}BiS_2$. We found that both As-grown and HP samples show bulk superconductivity. The lattice constant of *a* and *c* axis does not show a remarkable differences between the As-grown and HP samples. Nevertheless, the peak symmetry of the (200) peak seems to become more symmetric by high-pressure annealing while the (004) peak does not show such a difference, indicating that the crystal structure within the *ab* plane changed to a higher-symmetric phase, more perfect tetragonal, by high-pressure annealing in $CeO_{0.3}F_{0.7}BiS_2$. The $T_c$ of $CeO_{0.3}F_{0.7}BiS_2$ increased from $T_c^{zero}$ = 2.7 K to $T_c^{zero}$ = 3.7 K by high-pressure annealing. We assume that the increase in $T_c$ could be explained by a change in the crystal structure symmetry within the *ab* plane or optimization of the local atomic coordinates without any changes in lattice constants.

**Acknowledgements**

This work was partly supported by JSPS KAKENHI Grant Numbers 25707031, 26600077.

Figure captions

Fig. 1. X-ray diffraction profiles of the As-grown and the HP samples of $CeO_{0.3}F_{0.7}BiS_2$. (b) Enlarged X-ray profiles around the (004) peaks of the As-grown and the HP samples. (c) Enlarged X-ray profiles around the (200) peaks of the As-grown and the HP samples.

Fig. 2. Temperature dependences of electrical resistivity for the As-grown and the HP samples of $CeO_{0.3}F_{0.7}BiS_2$ below 15 K.

Fig. 3. (a) Temperature dependence of ZFC and FC magnetization for the As-grown $CeO_{0.3}F_{0.7}BiS_2$ sample. (b) Temperature dependence of ZFC and FC magnetization for the HP sample of $CeO_{0.3}F_{0.7}BiS_2$.

Figures

Fig. 1.

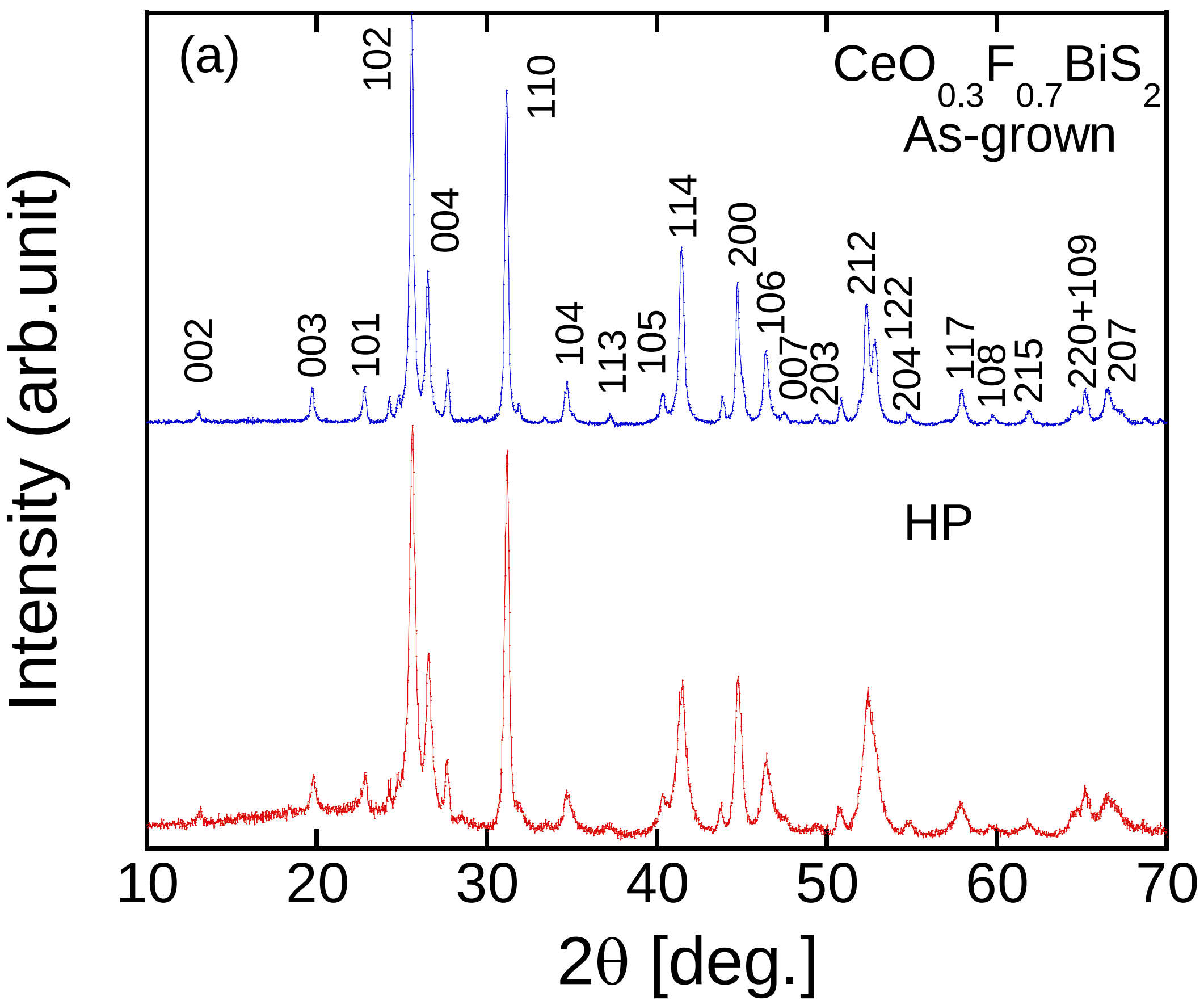

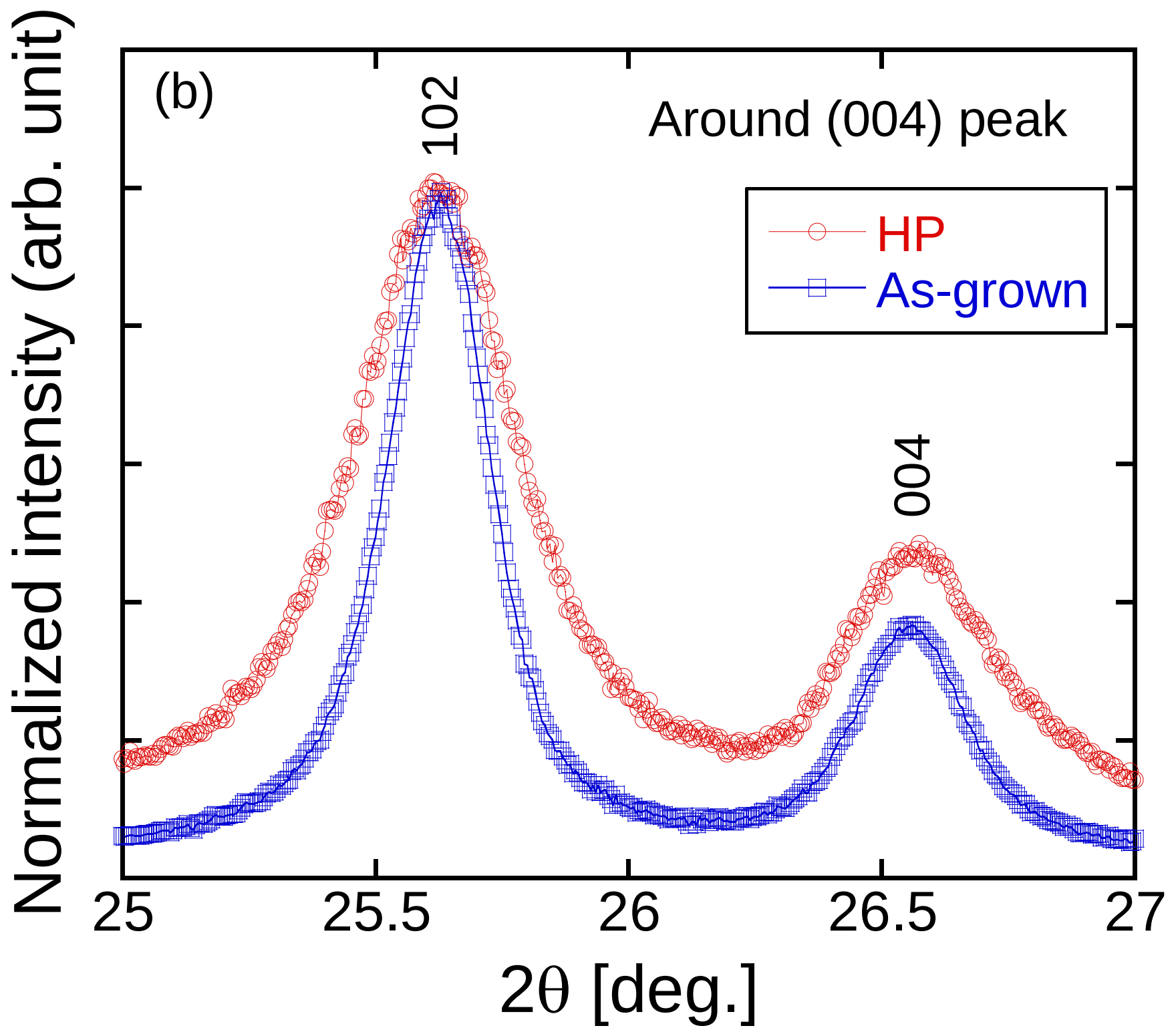

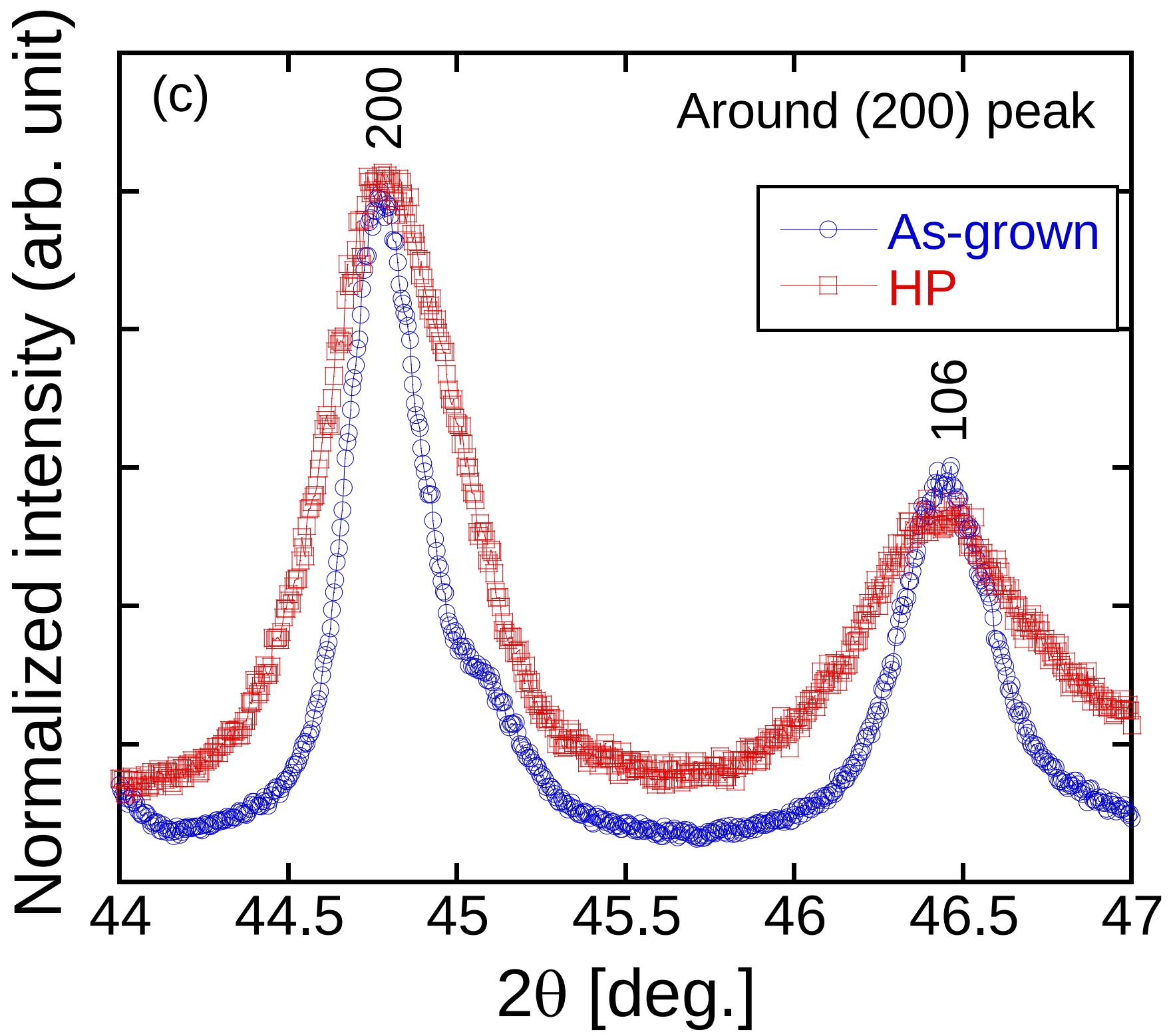

Fig. 2

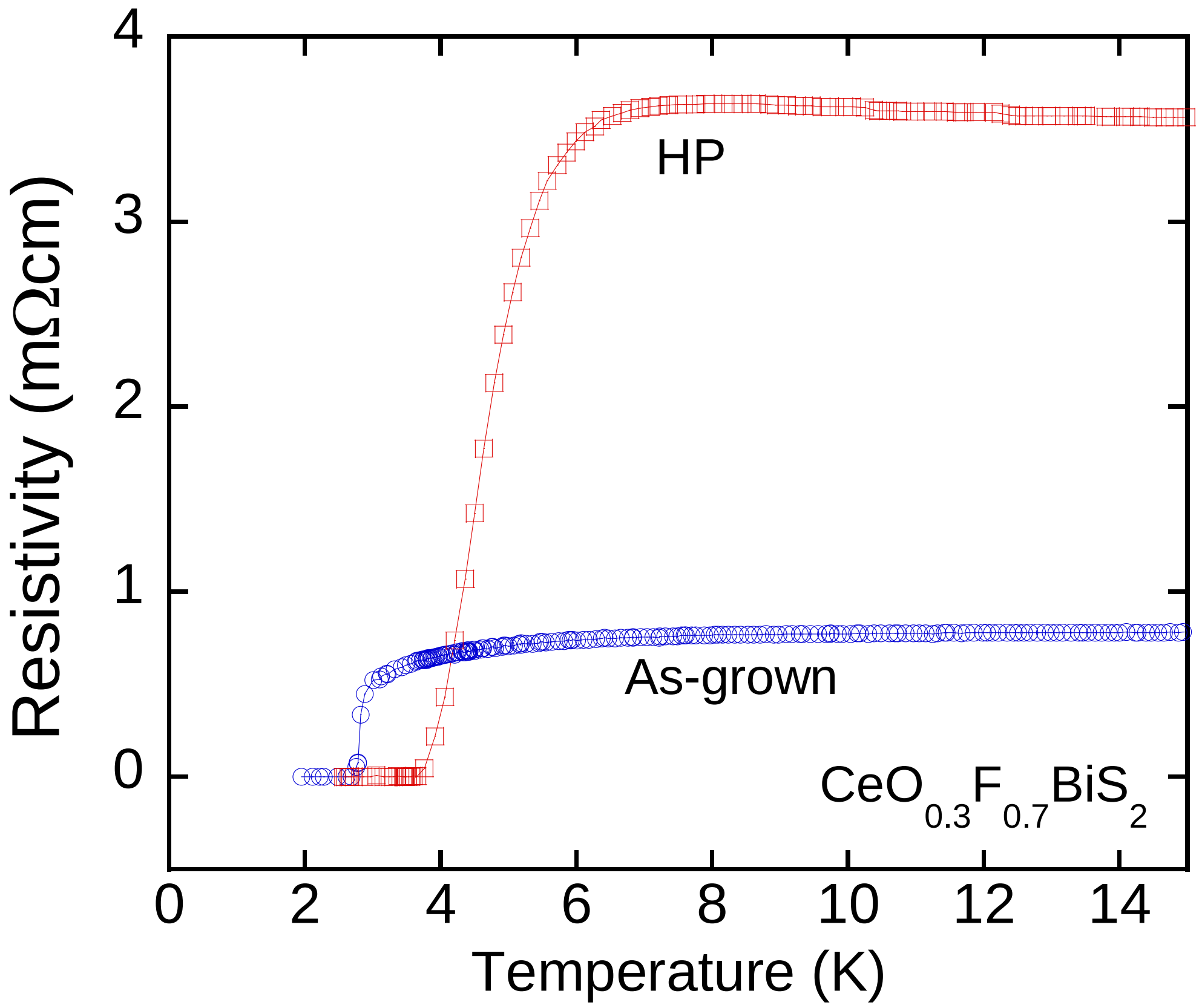

Fig. 3

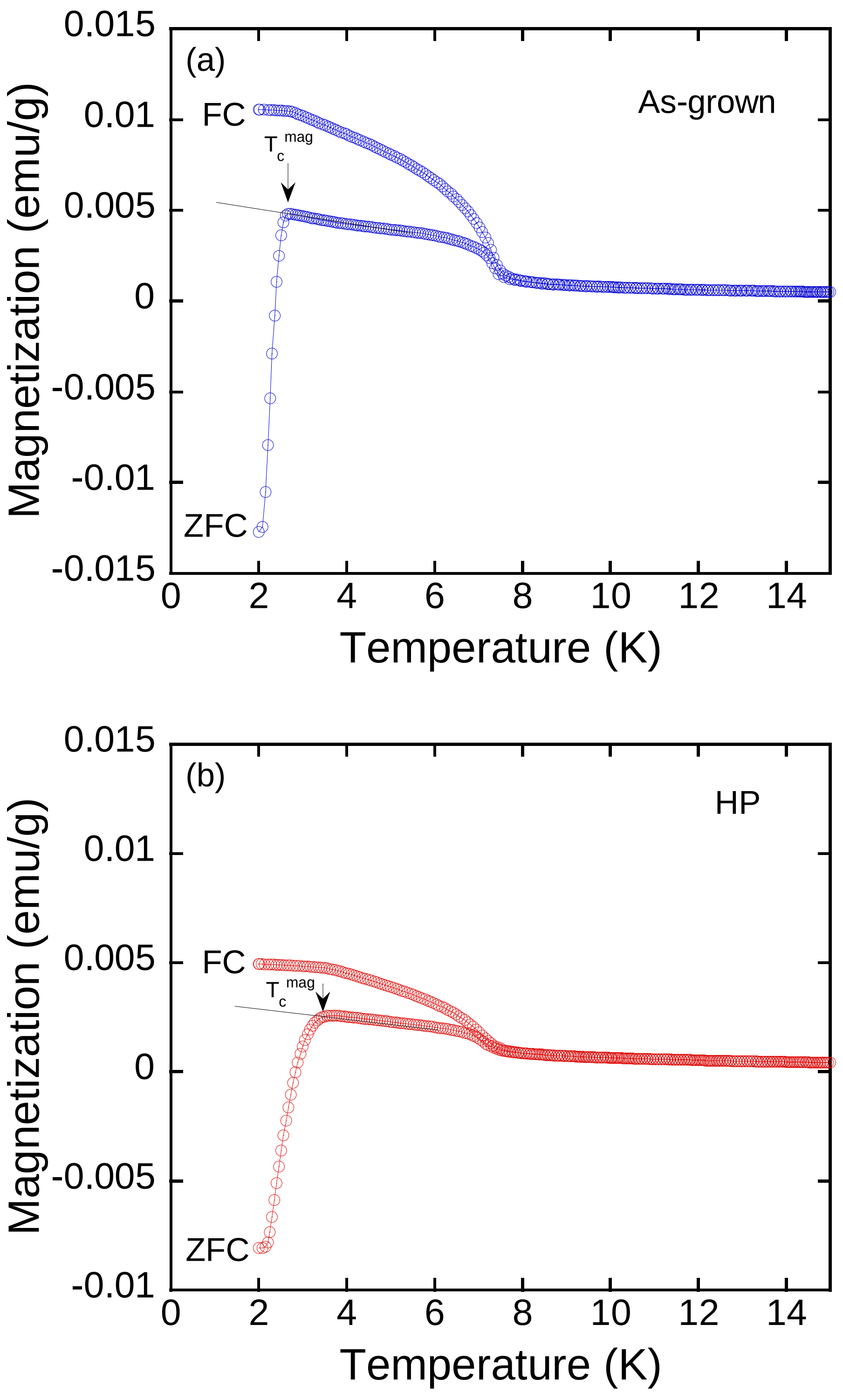